# Optical phased array using phase-controlled optical frequency comb

TAKASHI KATO[1, 2*] AND KAORU MINOSHIMA[1]

[1]*The University of Electro-Communications (UEC), 1-5-1 Chofugaoka, Chofu, Tokyo, Japan*
[2]*PRESTO, JST, 1-5-1 Chofugaoka, Chofu, Tokyo, Japan*
*\*takashi.kato@uec.ac.jp*

**Abstract:** We developed an optical phased array using an optical frequency comb and demonstrated its proof-of-principle. Optical phased arrays have been actively developed in recent years as a technology that can control the wavefront of light without any mechanical devices like phased array radar. Conventional optical phased arrays have been implemented using optical integrated circuits, but it has been difficult to achieve broadband operation with simple control. This is because control and calibration of a large number of phase modulators are required for each wavelength, and the dispersion of the waveguide makes whole bandwidth phase control of ultrashort pulses difficult. In contrast, we have developed a novel optical phased array that realizes wavefront control of ultrashort pulses generated by mode-locked laser by phase control of the comb, using high controllability of the comb and an optical array antenna with free-space optics. This is achieved by simply controlling the ratio of the two radio frequencies of the comb to realize a broadband optical phased array while suppressing environmental fluctuations. Experiments demonstrated broadband optical dot scanning at an optical frequency by forming an optical dot pattern and suppressing the environmental fluctuation by controlling the comb frequency. This innovative optical technology enables direct control of wavefronts by optical frequencies, i.e. controlling transverse modes by longitudinal modes.

## 1. Introduction

### 1.1 Optical phased array

In recent years, optical phased arrays (OPAs) have been developed as a light source for 3D Lidar and laser projectors [1]. Phased array radar, which are widely used in the radio wave domain, are a technology that controls the wavefront of radio waves by emitting phase-controlled radio waves from each element of an antenna array, and optical phased array is an extension of this technology to the optical domain.

The principle of OPA is similar to that of phased array radar and can be achieved by emitting phase-modulated light from an optical array antenna. First, a CW laser is split and introduced into a phase modulator, which gives it a designed relative phase amount. These beams are then emitted from the optical array antenna, where the individual wavefronts interfere with each other and form an interference wavefront according to the given phase pattern. By controlling the phase amount, the light intensity can be spatially isolated, and operations such as an optical point scanning are possible. By following this process in reverse, the system can also function as a receiver, and it is possible in principle to measure the wavefront of the incident light from the phase difference detected by the individual optical antennas. From the above functions, optical phased arrays are a wavefront control technology capable of controlling and measuring the wavefront of light.

With the recent development of optical integrated circuits, research and development of optical phased arrays implemented by the optical waveguide is currently active. The basic principle of operation is the method shown in Fig. 1. Research groups around the world are competing to find out how to implement more phase controllers and optical antennas. Examples have already been reported of the implementation of approximately 1000 optical antennas [2] and the optical integrated circuit consisted of OPA and detection system [3], and implementation in the near-infrared [1], visible light [4] and mid-infrared region [5], where research has been particularly active in recent years, is also underway. In addition, by incorporating a function as a lidar, a ranging method, applications as a high-speed 3D lidar without any mechanical drive such as galvano-mirrors are being promoted [2].

As described above, OPAs, which can realize high-speed optical point scanning by eliminating any mechanical drive, are expected to have high potential as a light source for optical measurement, but there are several issues to be addressed.

First, it is difficult to use a broadband ultrashort pulse light source, i.e. mode-locked laser, as a light source for an OPA with an optical waveguide. Phase control is achieved using electro-optical modulators (EOMs) and temperature-controlled optical path length variation, but it is difficult to accurately control the phase pattern of each optical antenna over the whole bandwidth due to wavelength dependence and dispersion changes. In addition, the dispersion of the optical waveguide gives a chirp to the ultrashort pulses travelling along the optical waveguide, which changes the pulse width. In particular, the phase control part for constructing a phased array and the optical path of the optical antenna are independent, and if the length of each optical waveguide varies even slightly, the phase of the ultrashort pulses emitted from the optical antenna will differ according to wavelength.

Second, the number of control circuit is increased by increasing the optical antennas. In the case of waveguide OPAs, precise control is required for the optical path length of the many implemented phase control parts and the entire waveguide. For example, the phase of propagating light in an EOM made of lithium niobate changes according to the voltage applied. However, as many EOMs are mounted independently, the required voltage for a certain amount of phase varies depending on the individual EOMs. When the phase amount is changed by changing the optical path length through local temperature control, it is difficult to stabilize the phase amount because delicate temperature adjustment is required. In addition, because OPAs use interference, the interferometer is used from the point where the light source is branched into many branches to the point where the wavefront is formed. Therefore, temperature control of the waveguide substrate is also important, as changes in optical path length due to changes in the temperature of the entire optical integrated circuit directly affect the wavefront to be formed. As these parameters need to be accurately controlled, the electrical circuits for control increase with the number of optical antennas, making accurate wavefront control difficult.

Although the use of CW light sources is useful enough for applications, the use of ultrashort pulses can provide a more powerful optical technology. Femtosecond ultrashort pulses are a light source whose usefulness has been demonstrated in various optical measurement fields, such as distance measurement and spectroscopy, and if the optical frequency comb described below is used as a light source, wide dynamic range measurement, which is difficult to achieve using existing methods, can be achieved. Therefore, if there is an OPA that can adapt to ultrashort pulse light sources, it is possible to adapt a highly functional optical measurement method to each point of a two-dimensional surface, and because an OPA can be controlled at high speed, it is also possible to achieve both high spatial resolution and measurement speed. Therefore, this paper proposed an OPA based on a completely new principle that is adaptable to ultrashort pulses.

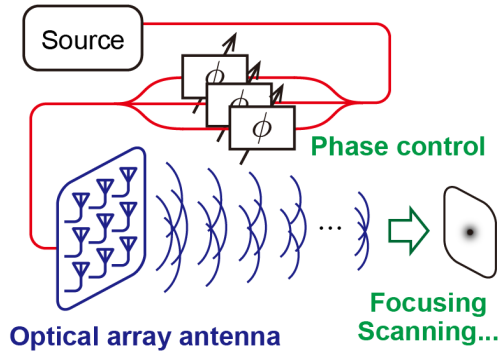

Fig. 1. The principle of optical phased arrays

*1.2 Optical frequency comb*

An optical frequency comb (OFC) is a phase-controlled mode-locked laser. In the frequency domain, it constitutes a group of spectra equally spaced like a comb, while in the time domain, an OFC forms ultrashort pulse trains on the order of femtoseconds. The optical frequency (THz to PHz) can then be freely controlled by two radio frequencies (MHz to GHz), the repetition frequency $f_{rep}$ and the carrier envelope frequency $f_{ceo}$, and a frequency stability of more than 15 orders of magnitude can be achieved by stabilization with a microwave frequency reference, e.g. atomic clock. Each frequency can directly control the characteristics of the ultrashort pulse train in the time domain, with controlling the pulse interval by $f_{rep}$ and the phase of the ultrashort pulses over the whole bandwidth by $f_{ceo}$. As described above, OFCs can freely control the time and frequency domain of light, and their extremely high stability can contribute to reducing the measurement uncertainty of optical measurements. In fact, they dramatically improve the accuracy and dynamic range of distance measurements, and their usefulness has been demonstrated in a number of previous studies [6-13]. With the development of OPA using OFCs, applicability of comb distance measurement techniques can be further expanded. In the OPA of this study, the ability to freely control the pulse interval and the phase of the whole bandwidth is actively exploited.

## 2. Methods

The OPA proposed in this paper is realized by controlling the carrier phase of ultrashort pulses generated by an OFC over the whole bandwidth by means of $f_{rep}$ and $f_{ceo}$, and by constructing an optical antenna with a cavity-type optical delay line. A conceptual diagram of the proposed OPA with OFC is shown in Fig. 2.

First, the carrier phase of the ultrashort pulses was controlled to generate a pulse train with controlled carrier phase difference of pulses. By controlling the frequency ratio of the $f_{rep}$ and $f_{ceo}$ of the OFC, whole bandwidth phase control of the ultrashort pulse train can be achieved. For example, if the frequency of the OFC is maintained such that $f_{rep} = 4f_{ceo}$, an ultrashort pulse train is output with the carrier phase changing by 90°. With this control method, we have developed a one-shot 3D imaging method using an optical processing technique based on an OFC, which is already operating at a practical level [14, 15]. This means that the phase relationship of the individual pulses can be freely controlled by adjusting this frequency ratio.

The phase-controlled ultrashort pulse train is then introduced into the optical antenna. OPAs emit pulses with a constant phase difference from different positions, and the interference between them produces an arbitrary wavefront. The optical antenna used in this method spatially re-arranged the phase-shifted pulse trains. As shown in Fig. 2, in this method, the ultrashort pulse trains are re-arranged on a circumference by a multipath cavity (MPC) and the pulse interval and a cavity length of the MPC are matched so that a series of phase shifted pulses are simultaneously emitted from the MPC. This emission point corresponds to an optical antenna. A neutral density filter was used to suppress differences in the intensity of the individual pulses emitted at this time.

There are two key points to this optical antenna: first, it uses free-space optics, so there is no dispersion effect. Although the pulses pass through the glass on their way to the last emitting surface, all pulses emitted from each optical antenna will pass through the glass only once, so they are all given the same dispersion. This results in the output of ultrashort pulses with the desired phase relationship in a whole bandwidth, forming the interference wavefront designed in far field.

Another point is that the $f_{rep}$ can suppress fluctuation of the cavity length of the MPC, which fluctuates due to environmental disturbances: by setting the cavity length of the MPC to half of the pulse interval, the pulses of the OFC output from the MPC can interfere with each other. However, because it is a free-space optical setup, if the cavity length fluctuates, the interference between pulses fluctuates and the wavefront formed by the OPA becomes unstable. Therefore, by locking the $f_{rep}$ to the cavity length of the MPC that has fluctuated due to environmental disturbance, the pulse-to-pulse interference of the OFC output from the MPC can be stabilized. This enables robust OPA without any mechanical drives such as delay stages.

To demonstrate the proof-of-principle of the method, a dot pattern was first generated using only three optical antennas, and the environmental disturbance of the MPC was suppressed by frequency control of the OFC. Next, the optical dots were further isolated by five optical antennas, and the optical dot scanning was demonstrated by frequency control of the OFC. Finally, a variable bandpass filter (BPF) was inserted in the optical setup to verify the broadband operation of the proposed method.

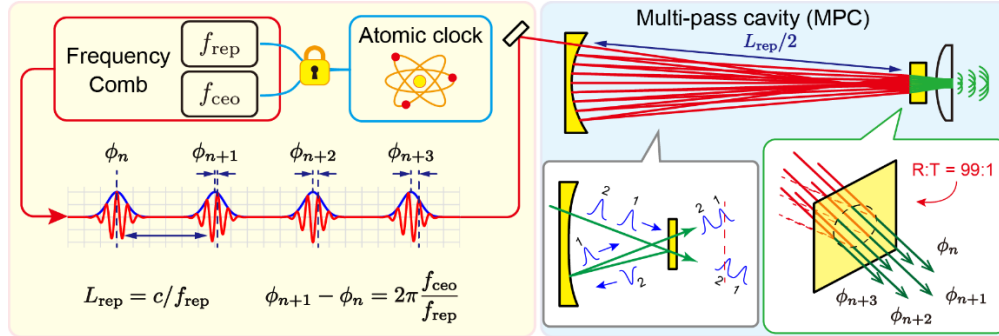

Fig. 2 A principle of the proposed OPA with OFC

## 3. Experiments

The experimental system is shown in Fig. 3(a). An Er-doped fiber laser ($f_{rep}$: 108 MHz, center wavelength: 1.55 μm, bandwidth: 21.70 nm) was used as the optical comb light source and a 1f-2f interferometer was constructed to control $f_{rep}$ and $f_{ceo}$. A function generator was used to control the constant frequency, and the frequency stability was approximately $10^{-11}$ (gate time: 1 s). The phase-controlled OFC was amplified using an Er-doped fiber amplifier (EDFA) and a pulse train of 113 mW was introduced into the MPC. The OFC's spectrum and the temporal pulse shape measured by autocorrelator are shown in Fig. 3(b). The spectral waveform was soliton-like and the pulse width after EDFA was 1.21 ps. The dispersion of the pulse induced into the MPC here was not adjusted, so it was not a Fourier transform limited pulse; after MPC output, the controlled interference wavefront was captured by the InGaAs camera through the wedge prism optics.

The detail of MPC is shown in Fig. 4. The MPC in Fig. 3(a) consisted of a concave mirror with a radius of curvature of 1.4 m and a plane mirror. The two mirrors were coated with Au. The plane mirror was polished on the back surface and a phase difference pulse of about 1 % was output from the back surface of the plane mirror. This means that the reflection points of the plane mirror worked as an optical antenna. In order to make each pulse output from the MPC's plane mirror interfere with each other in far field, the optical axis was tilted using wedge prisms. The ray trace of the MPC is shown in Fig. 4(a). The red and blue points are the reflection points on the concave and plane mirrors respectively, and the MPC has 17 optical antennas. Only three and five of these antennas were used in the experiment. The detailed setup of the MPC is shown in Fig. 4(b), where the MPC was built on a breadboard made of low thermal expansion material (SiAlON) to reduce environmental disturbance. Due to the size limitation of the breadboard, a V-shaped cavity was constructed using one plane gold mirror.

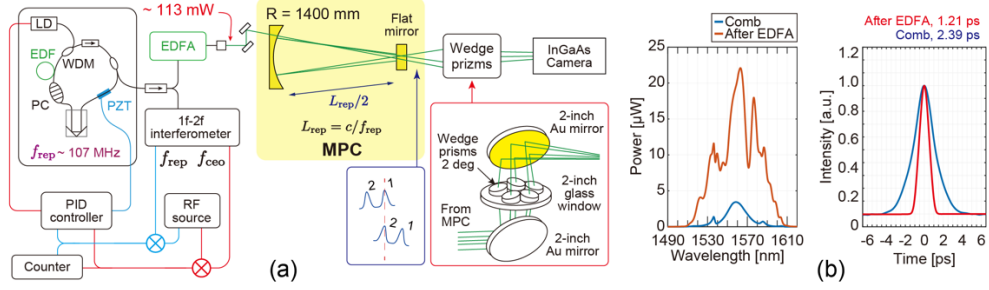

Fig. 3 (a)Experimental setup (b)Spectrum and temporal pulse shape of OFC

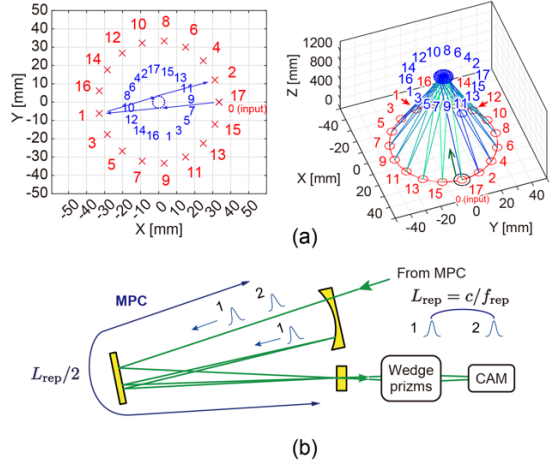

Fig. 4 (a)Ray trace of the MPC (b)Detailed setup of the MPC

## 4. Results

### 4.1 Optical dot pattern formed by three optical antennas and stabilization by controlling $f_{rep}$

The dot pattern formed by OPA using three optical antennas is shown in Fig. 5(a). Three pulses are interfering simultaneously on the time domain, and as the wavenumber vector of each pulse was different, the interference wavefront formed a dot pattern as shown in Fig. 5(a). This dot pattern can be controlled by the relative phase difference between each pulse, which can be controlled by the ratio of $f_{rep}$ and $f_{ceo}$ of the OFC. On the other hand, fluctuations of the optical path length of the MPC due to environmental disturbance can similarly change the interference wavefront. To suppress this fluctuation, it is necessary to synchronize the pulse interval of the OFC with the fluctuating cavity length of the MPC. Therefore, the pulse interval of the OFC was synchronized with the cavity length of the MPC by controlling $f_{rep}$ using the interference signals of the two optical antennas.

The variation of the interference signal at one point on the interference wavefront is plotted in Fig. 5(b). Initially, the interference signal changed significantly due to the optical path length variation of the MPC, but when the fluctuation was suppressed by the $f_{rep}$ control, the interference signal became approximately constant. This result shows that robust OPA can be achieved by frequency control of the OFC, even if the optical antenna is a spatial optical setup.

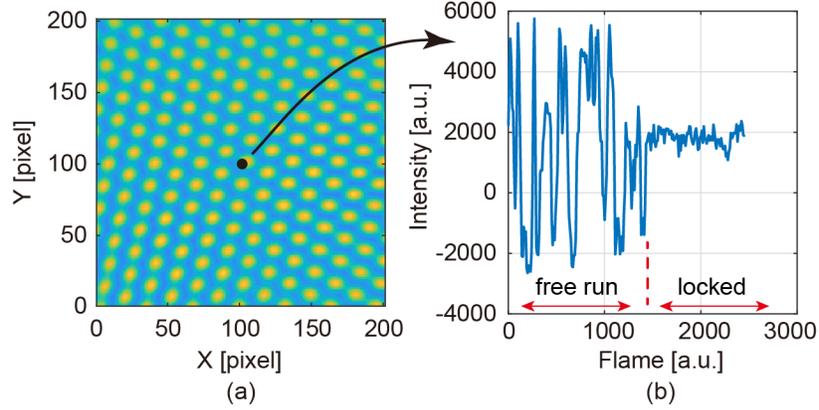

Fig. 5 (a)Optical dot pattern formed by three optical antennas
(b)Intensity at one pixel, (X, Y) = (104, 104) on (a), with and without stabilization controlling $f_{rep}$.

## *4.2 Localization of an optical dot formed by five optical antennas and optical dot scanning by controlling the frequency ratio*

Next, the number of optical antennas was increased to five and a dot pattern was constructed with a reduced number of optical dots. The dot pattern with five optical antennas is shown in Fig. 6(a). Five interference fringe images are shown in Fig. 6(a) when only two of the five pulses interfered with each other. They show different interference fringes because the wavenumber vector of each pulse and the phase difference between the pulses were different. As they all interfered at the same timing, when combined, these formed the dot pattern with fewer optical dots than in Fig. 5(a).

Focusing on one optical dot in Fig. 6(a), the optical dot when the frequency ratio of the OFC was varied are shown in Fig. 6(b). Images and 3D plots for frequency ratios ($f_{rep}$ / $f_{ceo}$) of 4.078 and 7.518 are shown in Figs. 6(b-1) and 6(b-2). The $f_{rep}$ was fixed at $f_{rep}$ = 107.639749 MHz and varied by $f_{ceo}$ only. Environmental disturbances stabilized the experimental setup to a negligible change. As Fig. 6(b) shows, the position of the optical dot moved with the frequency ratio, and the intensity between the optical dot and the rest were about 10 times different. This result indicates that the optical dot's position can be controlled by the frequency ratio of the OFC.

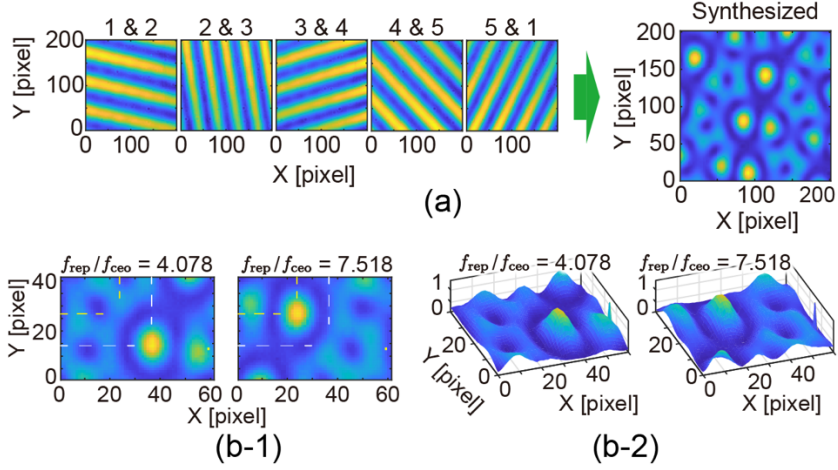

Fig. 6 (a)Optical dots formed by five optical antennas. (b-1) and (b-2) show images and 3D plots of optical dots controlled by the frequency ratio.

*4.3 Broadband operation*

Finally, the broadband operation of the proposed method was evaluated. In the experimental system shown in Fig. 3(a), a variable BPF (center wavelength: 1530 ~ 1610 nm, bandwidth: 3 nm) was inserted just before the MPC. This created an interference wavefront using only specific wavelengths; the images of dot patterns taken with changing center wavelengths of 1530, 1555, 1580 and 1610 nm are shown in Fig. 7. The intensity of each image was normalized so that the noise intensity of each image was different. This is due to the fact that the spectral intensity of the OFC was not uniform. During the wavelength sweep, the control of the OFC and the alignment of the MPC were not changed in any way, but the proposed OPA succeeded to form similar dot patterns at each wavelength.

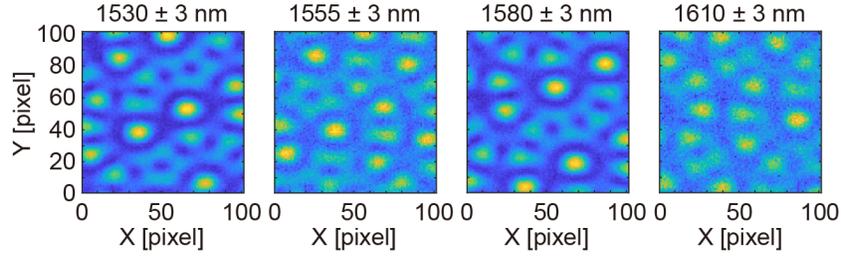

Fig. 7 Normalized images of optical dot patterns with changing wavelength from 1530 to 1610 nm

## 5. Discussions

This experimental setup used the MPC with a cavity length of approximately 1.4 m, which depended on the $f_{rep}$ of the OFC. Since this method requires the cavity length to be half of the pulse interval, OPA can be realized with a more compact MPC if the OFC with highly $f_{rep}$ is used. For example, using an electro-optic frequency comb, EO-comb, with a $f_{rep}$ of 10 GHz or a microcomb with a $f_{rep}$ of 100 GHz, an OPA can be constructed using an MPC with a cavity length of 15 mm and 1.5 mm, respectively. In particular, if the microcomb is used as a light source, it is possible to integrate everything from the light source to the MPC into a chip size.

In the experiment, five optical antennas were used to construct the interference wavefront, but as the number of optical antennas is increased, fewer optical points will remain on the image, and if the number of points is sufficiently large, only one point will remain. In particular, the optical antennas in this method were arranged in a circumferential configuration. If 32 optical antennas in a circumferential configuration can be constructed and generate the optical dot, an OPA constructed by 256 optical antennas arranged in a gridded configuration was required to realize the same sharpness of the optical dot [16]. So, the OPA constructed by few optical antennas arranged in a circumferential configuration can be applied to practical applications.

In Fig. 7, the position of the optical dots changes when the interference wavefront was constructed with different wavelengths, because the wavenumber vector of the pulses emitted from the five optical antennas changed with wavelength. This characteristic can be used for two-dimensional scanning of the optical dot [17]. In the conventional method using optical integrated circuits, it was necessary to change the control amount of the phase modulation part when the wavelength was changed. However, in this method, the phase between pulses emitted from each optical antenna was controlled over the whole bandwidth by the frequency ratio of the OFC, and because the optical antenna was constructed using free-space optics, the interference wavefront over the whole bandwidth can be stably controlled. This is a useful feature for various applications using the OPA.

## 6. Conclusion

Using a phase-controlled OFC, we proposed and demonstrated a novel OPA method that enables wavefront control of ultrashort pulses across the entire spectral bandwidth. The proposed OPA realized to control broadband wavefront by optical frequency with high stability, repeatability and speed. Because the proposed method uses free-space optics as an optical antenna, it applies to broad wavelength regions, challenging to access through conventional methods. Increasing the number of optical antennas can generate more apparent focus spots and higher-order transverse modes. Controllable tight focus spots can directly induce nonlinear optical effects at the target point in the sample, making them particularly appealing for multiphoton imaging. In addition, combining the high accuracy of OFC with a long-range distance measurement method could enable high-precision and high-speed 3D Lidar.

**Funding.** Japan Science and Technology Agency (JST), PRESTO (JPMJPR2005).